\documentclass[a4paper]{jpconf}
\usepackage{graphicx}
\usepackage{latexsym}
\usepackage{iopams}

\newcommand{\Ahat}{{\hat A}}
\newcommand{\Bhat}{{\hat B}}

\newcommand{\diag}{{\rm diag}}

\begin{document}
\title{Entanglement generation in relativistic cavity motion}

\author{David Edward Bruschi$^{1,2}$, 
Jorma Louko$^1$ 
and 
D.~Faccio$^3$}

\address{$^1$
School of Mathematical Sciences, 
University of Nottingham, 
Nottingham NG7 2RD, 
United Kingdom}
\address{$^2$
School of Electronic and Electrical Engineering, 
University of Leeds, 
Leeds LS2 9JT, 
United Kingdom}
\address{$^3$
School of Engineering and Physical Sciences, 
David Brewster Building, 
Heriot-Watt University, SUPA, Edinburgh
EH14 4AS, United Kingdom}

\address{$\phantom{xxx}$}                            
\address{January 2013\footnote[4]{Based on a talk    
given by JL at DICE2012                              
(Castiglioncello, Italy, 17--21 September 2012).}}   


\begin{abstract}
We analyse particle creation and mode mixing for a quantum
field in an accelerated cavity, assuming small accelerations but 
allowing arbitrary velocities, travel times and travel distances, 
and in particular including the regime of relativistic velocities. 
As an application, we identify a desktop experimental scenario where
the mode mixing resonance frequency in linear sinusoidal motion or in uniform 
circular motion is significantly below the
particle creation resonance frequencies of the Dynamical Casimir Effect, 
and arguably at the threshold of current technology. 
The mode mixing acts as a beamsplitter quantum gate,
experimentally detectable not only via fluxes or particle numbers 
but also via entanglement generation. 
\end{abstract}


\section{Introduction and overview}

As anyone caught by a police speed camera can testify, spacetime
kinematics affects photons that are present in spacetime. For a
relativistic quantum field, spacetime kinematics can also create
particles where initially there were none, and the very notion of a
``particle'' depends on the motion of an observer. In flat spacetime,
celebrated examples are the thermality seen in Minkowski vacuum by
uniformly accelerated observers, known as the Unruh
effect~\cite{unruh,Crispino:2007eb}, and the creation of particles by
moving boundaries, known as the Dynamical {(or non-stationary)}
Casimir Effect (DCE)~\cite{Dodonov:advchemphys,Dodonov:2010zza}.  In
curved spacetime, a celebrated example is the Hawking radiation
emitted by black holes~\cite{hawking}.

There has been significant recent interest in harnessing consequences
of spacetime kinematics to serve quantum information tasks. The broad
issue is how entanglement and the fidelity of quantum communication
are affected by observer motion and by spacetime curvature
\cite{Alsing:2003es,FuentesSchuller:2004xp,Bruschi:2010mc,Downes:2010tv,Bruschi:2012vb,Bruschi:2011ug,Friis:2011yd,downes-ralph-walk:comm}.
A~specific question is how entanglement may be created and used to
construct quantum gates
\cite{Friis:2012tb,Bruschi:2012uf,Friis:2012nb,Friis:2012ki}.
A~recent survey can be found in~\cite{Rideout:2012jb}.

In this contribution we consider a quantum field in an accelerated,
perfectly confining cavity in Minkowski spacetime.  As the cavity
walls shield the field from acceleration horizons, the cavity field
cannot be expected to exhibit thermality via the Unruh effect when the
acceleration is linear and
uniform~\cite{schutzhold-unruh:comment-tele}.  While the absence of
the conventional Unruh effect can hence be regarded as a disadvantage
of a cavity field, the field will nevertheless respond to the cavity's
acceleration in a nontrivial fashion, and the confinement has
technical advantages in that a discrete spectrum makes a number of
entanglement quantifiers mathematically well
defined~\cite{plenio-virmani:review} and conceptual advantages related
to localisation issues in relativistic quantum measurement
theory~\cite{sorkin-impossible,Benincasa:2012rb}.  Finally, a
practical advantage of a cavity field is that the cavity can be given
a variety of travel scenarios of interest, including periodic shakes
that appear in the DCE
literature~\cite{Dodonov:advchemphys,Dodonov:2010zza}.

In Sections \ref{sec:pert1+1} and \ref{sec:pert3+1} we review a
recently-developed formalism \cite{Bruschi:2012pd} for a relativistic
field in a moving cavity that is assumed mechanically rigid, in the
sense that the cavity's shape in its instantaneous rest frame remains
unchanged even though the velocity may change in time.  The
acceleration is assumed to be small compared with $c^2$ divided by the
cavity's size but may otherwise have arbitrary time dependence, both
in magnitude and in direction. The velocities, travel times and travel
distances are unrestricted, so that the treatment remains valid even
when the velocities become relativistic. To linear order in the
acceleration, the Bogoliubov coefficients between inertial motions at
early and late times can then be expressed in terms of the Fourier
transform of the cavity's acceleration.

In Section \ref{sec:periodic} we apply the formalism to two periodic
motions, a linear sinusoidal motion and a uniform circular
motion~\cite{Bruschi:2012pd}. We identify a configuration that brings
the mode mixing resonance frequency in these motions into an
apparently experimentally feasible regime, significantly below the
experimentally prohibitive particle creation resonance frequencies
known from the DCE
literature~\cite{Dodonov:advchemphys,Dodonov:2010zza}.
A~geometrically optimised desktop experimental scenario at optical
wavelengths is seen to produce significant mode mixing within a
millisecond of shaking at megahertz frequencies. From the quantum
information viewpoint the mode mixing acts as a beamsplitter quantum
gate. The experimental verification opportunities can hence utilise
entanglement phenomena, in addition to conventional observations of
fluxes or particle numbers.

We conclude in Section \ref{sec:conclusions} by summarising the
results and discussing the prospects to bridge the gap to an actual
experiment that would create entanglement by mechanical motion on the
desktop.

We set from now on $c=\hbar=1$, reinstating units by dimensional analysis where appropriate.

\section{\label{sec:pert1+1}Rigid cavity at small accelerations: (1+1) Minkowski}

\subsection{Setup}

We start with a cavity in $(1+1)$-dimensional Minkowski spacetime.
The cavity is assumed mechanically rigid, in the sense that it
maintains constant length $L$ in its instantaneous rest frame.  We
denote by $\tau$ the proper time at the centre of the cavity and by
$a(\tau)$ the proper acceleration at the centre of the cavity.  To
maintain rigidity, the acceleration must be bounded by 
$|a|L < 2$~\cite{Bruschi:2011ug}.

The cavity contains a real scalar field $\phi$ of mass~$\mu_0\ge0$,
with Dirichlet boundary conditions at the walls.

\begin{figure}[t!]
\begin{center}
\includegraphics[width=17pc]{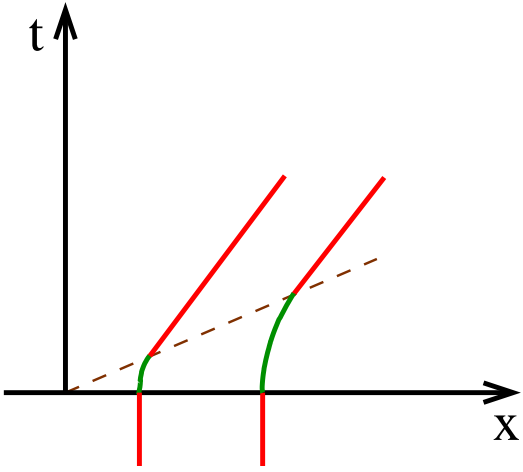}\hspace{3pc}%
\raisebox{1.5pc}[0pt][0pt]{\begin{minipage}[b]{16pc}\caption{\label{oneboxfig}A $(1+1)$ 
cavity worldtube 
in which the initial and final inertial segments are joined by one 
segment of uniform acceleration. 
During the accelerated segment the cavity is static 
with respect to the boost Killing vector
$x\partial_t + t\partial_x$, 
whereas during each of the inertial segments 
the cavity is static with respect 
to a Minkowski time translation Killing vector.}
\end{minipage}}
\end{center}
\end{figure}

We assume $a$ to vanish at early and late times but to be nonvanishing
at some intermediate times.  An example where $a$ is piecewise
constant with exactly two points of discontinuity is shown in
Figure~\ref{oneboxfig}.

The inertial segments of the cavity worldtube at early and late times
are static: the cavity is dragged along a Minkowski time translation
Killing vector. These timelike Killing vectors provide a distinguished
definition of positive and negative frequencies tailored to the
cavity's motion at early and late times, resulting to field modes with
the angular frequencies $\omega_n = \sqrt{\mu_0^2 + (\pi n /L)^2}$,
$n=1,2,\ldots$, and quantisation leads to the familiar Fock space of
an inertial cavity. However, because $a$ is nonvanishing at
intermediate times, the early time and late time vacua do not need to
coincide, nor do the early time and late time notions of
particles. The task is to find the Bogoliubov transformation that
relates the early time and late time Fock spaces.

\subsection{Solution}

When the accelerations are so small that $|a|L \ll1$, the problem has
a solution that builds on the following three key
observations~\cite{Bruschi:2011ug,Bruschi:2012pd}.

First, while a cavity with constant nonvanishing $a$ is not inertial,
it is nevertheless static: the cavity is dragged along a boost Killing
vector, as shown in Figure~\ref{oneboxfig}. The cavity can
be regarded as uniformly accelerated, even though the value of
the acceleration varies between spatial points within the cavity,
attaining the values $2a/(2 \pm a L)$ at the two walls and being equal
to $a$ only at the centre. The boost Killing vector is timelike and
provides now a distinguished definition of positive and negative
frequencies tailored to the cavity's motion. For $\mu_0>0$, the field
modes can be written down in terms of modified Bessel functions and
the angular frequencies are determined implicitly by a transcendental
equation. For $\mu_0=0$, the field modes and their angular frequencies
have elementary expressions~\cite{Bruschi:2011ug}. 

Second, the Bogoliubov transformation between an inertial segment and
a uniformly accelerated segment has a Maclaurin expansion in the
dimensionless parameter~$a L$, involving at each order only elementary
expressions, and for $|a| L \ll1$ the linear order terms provide
already a good approximation~\cite{Bruschi:2011ug}.  Cavity world
tubes for which $a$ is piecewise constant with a finite number of
discontinuities can hence be treated by composing this
inertial-to-uniform transformation, its inverse, and the free
evolution of the field modes during the uniformly-accelerated
segments~\cite{Bruschi:2011ug,Bruschi:2012uf}.  The prototype cavity
worldtube in which the initial and final inertial segments are joined
by a single segment of uniform acceleration is shown in
Figure~\ref{oneboxfig}.

Third, acceleration that is not necessarily piecewise constant can be 
handled by passing to the limit~\cite{Bruschi:2012pd}. 
The Bogoliubov coefficient matrices $\alpha$ and $\beta$ 
from the initial inertial segment 
to the final inertial segment have the expansions 
\begin{eqnarray}
\alpha &=& e^{i\boldsymbol{\omega} (\tau_1-\tau_0)} \bigl(1 + \Ahat + O(h^2) \bigr),
\label{eq:Ahat-def}
\\
\beta &=& e^{i\boldsymbol{\omega} (\tau_1-\tau_0)} \Bhat + O(h^2), 
\label{eq:Bhat-def}
\end{eqnarray}
where $h(\tau) := L a(\tau)$, 
$\boldsymbol{\omega} = 
\diag (\omega_1, \omega_2,\cdots )$, 
\begin{eqnarray}
\Ahat_{nn} &=& 0 
\ , 
\\
\Ahat_{mn} &=& 
- i 
\frac{\pi^2 m n \bigl(1-(-1)^{m+n}\bigr)} 
{L^4\left(\omega_m - \omega_n\right)^2
\sqrt{\omega_m \omega_n}}
\int_{\tau_0}^{\tau_1} e^{-i({\omega}_m - {\omega}_n) (\tau-\tau_0)} \, h(\tau) \, d\tau 
\ \ \ \hbox{for $m\ne n$}, 
\label{eq:Ahat-int-components}
\\
\Bhat_{mn} &=&
i 
\frac{\pi^2 m n \bigl(1 - (-1)^{m+n}\bigr)} 
{L^4\left(\omega_m + \omega_n\right)^2
\sqrt{\omega_m \omega_n}}
\int_{\tau_0}^{\tau_1} e^{-i({\omega}_m + {\omega}_n) (\tau-\tau_0)} \, h(\tau) \, d\tau 
\ , 
\label{eq:Bhat-int-components}
\end{eqnarray}
the initial inertial segment ends at $\tau=\tau_0$ 
and the final inertial segment begins at $\tau=\tau_1$. 
To linear order in~$h$, the Bogoliubov coefficients are hence obtained 
by just Fourier transforming the acceleration.

\subsection{Outcomes}

While the perturbative solution
(\ref{eq:Ahat-def})--(\ref{eq:Bhat-int-components}) for the Bogoliubov
coefficients assumes the acceleration to be so small that $|h| \ll1$
over the cavity's worldtube, the velocities, travel times and travel
distances remain unrestricted.  The solution remains in particular
valid even when the velocities grow relativistic.  In this sense our
perturbative treatment is complementary to the small distance
approximations often considered in the DCE
literature~\cite{Dodonov:advchemphys,Dodonov:2010zza}, while of course
overlapping in the common domain of validity.

$\Ahat_{mn}$ and $\Bhat_{mn}$ scale linearly in~$h$, but their
magnitudes depend also crucially on whether $h$ changes slowly or
rapidly compared with the oscillating integral kernels in
(\ref{eq:Ahat-int-components}) and~(\ref{eq:Bhat-int-components}). 
It is useful to consider the two extreme limits individually. 

One extreme is the limit of slowly-varying~$h$. In this limit
$\Ahat_{mn}$ and $\Bhat_{mn}$ vanish, as is seen from
(\ref{eq:Ahat-int-components}) and (\ref{eq:Bhat-int-components}) by
the Riemann-Lebesgue lemma, and as can also be argued on more general
adiabaticy grounds~\cite{schutzhold-unruh:comment-tele,unruh-private}.
The Bogoliubov coefficients indeed evolve by pure phases over any
segment in which $h$ is constant, not just to linear order in $h$ as
seen from (\ref{eq:Ahat-int-components}) and
(\ref{eq:Bhat-int-components}) but also nonperturbatively
in~$h$~\cite{Bruschi:2011ug,Bruschi:2012pd}.  Particles in the cavity
are hence created by \emph{changes\/} in the acceleration, rather than
by acceleration itself. The cavity is in this respect similar to a
single accelerating mirror, which radiates only when its acceleration
changes in time~\cite{Fulling:Davies:76}.

The other extreme is the case of piecewise constant~$h$.  In this case
the contributions to $\Ahat_{mn}$ and $\Bhat_{mn}$ come entirely from
the discontinuous jumps in~$h$.  While this extreme may not be
experimentally realisable by a material cavity, it can be simulated by
a cavity whose walls are static dc-SQUIDs undergoing electric
modulation that simulates mechanical
motion~\cite{wilson-etal,Friis:2012cx}.

When $h$ is sinusoidal and its angular frequency 
$\omega_r$ equals the angular frequency of the integral kernel in 
(\ref{eq:Ahat-int-components}) or~(\ref{eq:Bhat-int-components}), 
the corresponding Bogoliubov coefficient will grow linearly in 
the duration of the sinusoidal motion. 
These resonance conditions read 
\begin{eqnarray}
&&\Ahat_{mn}:
\hspace{1ex}
\omega_r = |\omega_m - \omega_n| 
\ , 
\label{eq:Ahat-lin-resonance}
\\
&&\Bhat_{mn}:
\hspace{1ex}
\omega_r = \omega_m + \omega_n 
\ , 
\label{eq:Bhat-lin-resonance}
\end{eqnarray}
where in each case $m-n$ needs to be odd 
in order for the coefficient
to be nonvanishing. 
The particle creation resonance 
(\ref{eq:Bhat-lin-resonance})
is well known in the DCE 
literature~\cite{Dodonov:advchemphys,Dodonov:2010zza,Reynaud1,Mundarain:1998kz,Crocce:2001zz,Crocce:2002hd,Ruser:2005xg,Yuce:2008tb,ISI:000172183100003,yuce-ozcak,sch-plunien-soff,dodonov-andreata:1999,dodo-dodo-mizrahi:2005}. 
The mode mixing resonance (\ref{eq:Ahat-lin-resonance}) is also known 
\cite{Mundarain:1998kz,Crocce:2001zz,Crocce:2002hd,Ruser:2005xg,Yuce:2008tb,ISI:000172183100003,yuce-ozcak} 
but seems to have received significant attention only in situations 
where it happens to coincide with a particle creation resonance. 

\section{\label{sec:pert3+1}Rigid cavity at small accelerations: (3+1) Minkowski}

The above $(1+1)$-dimensional analysis generalises immediately to a
rectangular cavity of edge lengths $(L_x,L_y,L_z)$
in $(3+1)$-dimensional Minkowski spacetime provided the acceleration
keeps pointing in one of the cavity's three principal directions.  The
quantum numbers in the transverse directions remain inert, just
contributing to the effective $(1+1)$-dimensional mass~$\mu_0$. Note
that $\mu_0$ is hence positive even when the $(3+1)$-dimensional field
in massless. 

It is further immediate to compose segments in which the acceleration
points in one of the cavity's principal directions, as seen in the
cavity's instantaneous rest frame, even if this direction differs from
segment to segment.

Acceleration of unrestricted magnitude and direction would require new
input regarding how the shape of the cavity responds to such
acceleration. To \emph{linear\/} order in the acceleration, however,
boosts in different spatial directions commute, and we can treat
acceleration as a vector superposition of accelerations in the three
principal directions in the cavity's instantaneous rest
frame~\cite{Bruschi:2012pd}.  For sinusoidal acceleration with angular
frequency~$\omega_r$, the resonance conditions
(\ref{eq:Ahat-lin-resonance}) and (\ref{eq:Bhat-lin-resonance})
generalise to
\begin{eqnarray}
&&
\Ahat_{mnp,m'n'p'}:
\hspace{1ex}
\omega_r = |\omega_{mnp} - \omega_{m'n'p'}|
\ , 
\label{eq:3+1resonance-Ahat}
\\
&&
\Bhat_{mnp,m'n'p'}: 
\hspace{1ex}
\omega_r = \omega_{mnp} + \omega_{m'n'p'}
\ , 
\label{eq:3+1resonance-Bhat}
\end{eqnarray}
where 
$
\omega_{mnp} = 
\sqrt{\mu^2 
+ (\pi m / L_x)^2 
+ (\pi n / L_y)^2 
+ (\pi p / L_z)^2}
$, 
$\mu$ is the mass of the field, 
the quantum numbers $(m,n,p)$ are positive integers, 
and the difference in the quantum number in the
direction of oscillation needs to be odd.

\section{\label{sec:periodic}Desktop mode mixing experiment with periodic motion}

\begin{figure}[t!]
\begin{center}
\includegraphics[height=15pc]{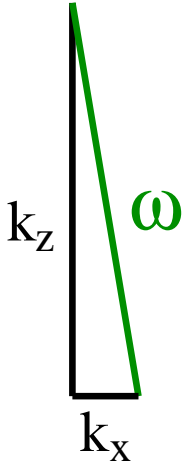}\hspace{5pc}%
\raisebox{1.5pc}[0pt][0pt]{\begin{minipage}[b]{20pc}\caption{\label{pythagorasfig}A momentum 
space configuration in which the
mode mixing resonance angular frequency
(\ref{eq:3+1resonance-Ahat}) is significantly lower than the angular 
frequencies of the individual cavity modes. 
The momenta of the quanta are aligned in the  
$z$-direction to a high degree of accuracy, so that 
$k_z \gg (k_x^2 + k_y^2)^{1/2}$.
When $k_x$ and $k_y$ change by the respective small amounts 
$\Delta k_x$ and~$\Delta k_y$, 
the changes in the angular frequency 
$\omega$ are
much smaller than $\Delta k_x$ and~$\Delta k_y$.}
\end{minipage}}
\end{center}
\end{figure}

The particle creation resonance angular frequency
(\ref{eq:3+1resonance-Bhat}) is comparable in magnitude to the angular
frequencies of the individual cavity modes and appears prohibitively
difficult to realise in experiments with mechanical
oscillations~\cite{Dodonov:advchemphys,Dodonov:2010zza}.  The mode
mixing resonance angular frequency (\ref{eq:3+1resonance-Ahat}) can
however be significantly lower when the cavity modes are highly
transverse to the direction of the oscillation, as illustrated in
Figure~\ref{pythagorasfig}.  We now outline a scenario that optimises
this lowering~\cite{Bruschi:2012pd}.

\subsection{Scenario}

We consider a massless field in $(3+1)$ dimensions and focus on quanta
whose wavelength $\lambda$ is much smaller than any of the
cavity's edge lengths.  The crucial input is to align the momenta
close to the $z$-direction, as shown in
Figure~\ref{pythagorasfig}, so that $\omega_{mnp} \approx 2\pi/\lambda
+ \frac14 \pi\lambda \bigl[ {(m/L_x)}^2 + {(n/L_y)}^2 \bigr] $. For
small changes in $m$ and $n$, the changes in $\omega_{mnp}$ are then
much smaller than the changes in the horizontal wave numbers $\pi m /
L_x$ and $\pi n / L_y$.

We let the cavity undergo linear or circular 
harmonic oscillation orthogonal to the $z$-direction, 
with amplitude $r_x$ ($r_y$) 
in the $x$-direction ($y$-direction). 
For motion in $x$-direction, 
the mode mixing resonance angular frequency 
(\ref{eq:3+1resonance-Ahat}) between modes $m$ and~$m'$, 
with $m-m'$ odd, is 
\begin{equation}
\omega_r
\approx 
{\textstyle\frac14} \pi \lambda L_x^{-2} \, \bigl|m^2 - {(m')}^2\bigr| \,,
\label{eq:resonance-x}
\end{equation}
and the growth rate of the mode mixing Bogoliubov coefficient is 
\begin{equation}
{\textstyle\frac{d}{d\tau}}|\Ahat_{\rm{res}}| 
\approx 
{\textstyle\frac12} \pi m m' r_x\lambda L_x^{-3}
\,. 
\label{eq:Ahatresabsdot-x}
\end{equation}
The lowest resonance
occurs for $m=1$ and $m'=2$. 
Similar formulas ensue for the $y$-resonance, and for circular motion 
both resonances are present. 

The experimental scenario starts by first trapping one or more quanta
in the cavity, in modes whose momenta are aligned close to the
$z$-direction.  The cavity is then made to oscillate perpendicularly
to the $z$-direction, linearly or circularly.  Finally, a measurement
on the quantum state of the cavity is performed.  The resonance mode
mixing that takes place during the acceleration is assumed to dominate
any effects due to the initial trapping and the final releasing of the
quanta.

\subsection{Desktop numbers}

For concrete numbers, we take $\lambda = 600\,$nm and 
$L_x = L_y= 1\,$cm (the value of $L_z$ does not need to be specified
provided $L_z \gg \lambda$). 
The lowest resonance angular frequency is 
$\omega_r 
\approx 4.2 \times 10^{6} \, {\rm{s}}^{-1}$, corresponding to an
oscillation frequency $0.7\,$MHz. With amplitude $1\,\mu$m, 
(\ref{eq:Ahatresabsdot-x}) and its $y$-counterpart show that the mode
mixing Bogoliubov coefficient grows to order unity within a
millisecond. Following the growth further would require theory beyond
our perturbative treatment. 

For linear motion, oscillation of micron amplitude at megahertz 
frequency may be achievable by using ultrasound to accelerate the cavity. 
Storing the quantum in the cavity for a millisecond could be challenging 
although recent advances indicate that it may be feasible~\cite{ultra-fabry-perot}.

For circular motion, the threshold angular velocity 
$\omega_r \approx 4.2 \times 10^{6} \, {\rm{s}}^{-1} 
\approx 4\times 10^7\,$rpm exceeds
the angular velocity $1.5 \times 10^5\,$rpm achieved by medical
ultracentrifuges~\cite{thermocientific}, although only by about two
orders of magnitude, and it is conceivable that this gap could be
bridged by a specifically designed system.

\section{\label{sec:conclusions}Conclusions and outlook}

We have quantised a scalar field in a rectangular cavity that is
accelerated arbitrarily in $(3+1)$-dimensional Minkowski spacetime, in
the limit of small accelerations but arbitrary velocities and travel
times. The Bogoliubov coefficients between inertial motion at early
and late times were expressed in terms of the Fourier transform of the
acceleration. For linear or circular periodic motions, we identified a
configuration in which the mode mixing resonance frequency is
significantly below the frequencies of the cavity modes.  

It can be verified that our scalar field analysis adapts in a
straightforward way to a Maxwell field with perfect conductor boundary
conditions: the two polarisation modes reduce to $(1+1)$ scalar fields
with respectively Dirichlet and Neumann boundary conditions, and the
Neumann condition Bogoliubov coefficients obey estimates that are
qualitatively similar to those given above~\cite{louko:maxwell}. The
mode mixing effects appear hence to be within the reach of a desktop
experiment with photons, achievable with current technology in its
mechanical aspects, if perhaps not yet in the storage capabilities
required of a mechanically oscillating optical cavity.

Assuming the cavity to be rectangular allowed us to treat the cavity
as spatially rigid in the small acceleration limit even when the
velocities may be relativistic and the direction of the acceleration
may vary in time.  We anticipate that the particle creation and mode
mixing effects are not qualitatively sensitive to the detailed shape
of the cavity, and this freedom could be utilised in the development
of a concrete laboratory implementation.

The experimental prospects could be further improved by filling the
cavity with a medium that slows light down~\cite{laupretre-slowlight}.
As the medium breaks Lorentz-invariance, our analysis would not be
directly applicable at relativistic velocities, but at nonrelativistic
velocities it may suffice to just include appropriate slowing-down
numerical factors in the angular frequencies.

Our experimental scenario involves significant mode mixing but no
significant particle creation.  One might therefore be tempted to
dismiss the scenario as a glorified speed camera effect and question
the interest of a laboratory observation.  However, mode mixing
without particle creation is known in quantum optics as a passive
transformation~\cite{simon-mukunda-dutta}, normally implemented by
passive optical elements such as beam splitters and phase plates, and
these transformations have a well understood capability to create and
degrade
entanglement~\cite{Friis:2012tb,Friis:2012nb,PhysRevLett.90.047904,Alsing:Fuentes:2012}.
For example, the mode mixing generates entanglement from an initial
Gaussian state only if this state is
squeezed~\cite{Friis:2012nb,PhysRevLett.90.047904,Alsing:Fuentes:2012}.
The oscillating cavity can hence be tuned to act as a beam splitter
quantum gate, creating or degrading entanglement in situations where
particles are initially present.  This means that the mode mixing
could be verified experimentally by observations of entanglement, and
the entangling power of the system could be potentially harnessed to
quantum information tasks.  We anticipate that observations of
entanglement will generally provide opportunities for experimental
verification of both particle creation and mode mixing effects that
are complementary to observations of fluxes or particle numbers.

\ack 
We thank Ivette Fuentes for the collaboration \cite{Bruschi:2012pd} on which 
this contribution is based. 
J.~L. thanks the organisers of DICE2012 
for the invitation to present this work. 
J.~L. acknowledges financial support from STFC [Theory Consolidated
Grant ST/J000388/1].
D.~F. acknowledges financial support from EPSRC project EP/J00443X/1.

\end{document}